\documentclass[preprint,superscriptaddress]{revtex4}

\usepackage{graphicx}
\usepackage{lipsum}
\usepackage{float}
\usepackage{amsmath}
\usepackage{amssymb}
\usepackage{mathtools}

\begin{document}

\title{Distance distribution in configuration model networks}

\author{Mor Nitzan}

\affiliation{Racah Institute of Physics, The Hebrew University, 
Jerusalem 91904, Israel}

\affiliation{Department of Microbiology and Molecular Genetics, 
Faculty of Medicine, The Hebrew University,
Jerusalem 91120, Israel}

\author{Eytan Katzav} 

\affiliation{Racah Institute of Physics, The Hebrew University, 
Jerusalem 91904, Israel}

\author{Reimer K\"uhn}

\affiliation{Department of Mathematics,
King's College London,
Strand, London WC2R 2LS, UK}

\author{Ofer Biham}

\affiliation{Racah Institute of Physics, The Hebrew University, 
Jerusalem 91904, Israel}

\begin{abstract}

We present analytical results for the distribution of shortest 
path lengths between random pairs of nodes
in configuration model networks.
The results, which are based on recursion equations, 
are shown to be in good agreement with numerical simulations 
for networks with 
degenerate, binomial and power-law 
degree distributions. 
The mean, mode and variance of the distribution
of shortest path lengths
are also evaluated. 
These results provide expressions for central 
measures and dispersion measures of the 
distribution of shortest path lengths
in terms of moments of the degree distribution,
illuminating the connection between the two distributions.
\end{abstract}

\maketitle

\section{Introduction}

The study of complex networks has attracted much attention in recent years.
It was found that network models provide a useful description of a large
number of processes which involve interacting objects
\cite{Barabasi2002,Caldarelli2007,Havlin2010,Newman2010,Estrada2011}.
In these models, the objects are represented by nodes and
the interactions are expressed by edges.
Pairs of adjacent nodes can affect each other directly.
However, the interactions between most pairs of nodes 
are indirect, mediated by intermediate nodes and edges.

A pair of nodes, $i$ and $j$, may be connected by a large number
of paths. The shortest among these paths are of particular 
importance because they are likely to provide the fastest 
and strongest interaction.
Therefore, it is of interest to study the distribution of shortest 
path lengths (DSPL) 
between pairs of nodes in different types of networks.
Such distributions, 
which are also referred to as distance distributions,
are expected to depend on the network 
structure and size.
They are
of great importance for the 
temporal evolution of dynamical processes on networks, 
such as signal propagation
\cite{Maayan2005}, 
navigation 
\cite{Dijkstra1959,Delling2009,Abraham2013}
and epidemic spreading 
\cite{Satorras2001,Satorras2015}.
Central measures of the DSPL such as 
the average distance 
between pairs of nodes,
and extremal measures such as
the diameter 
were studied
\cite{Bollobas2001,Watts1998,Fronczak2004}.
However, apart from a few studies
\cite{Newman2001,Blondel2007,Dorogotsev2003,Hofstad2007,Hofstad2008,Esker2008},
the entire DSPL
has attracted 
little attention.

Recently, an analytical approach was developed for calculating 
the DSPL 
\cite{Katzav2015}
in the
Erd{\H o}s-R\'enyi (ER) network,
which is the simplest mathematical model of a random network
\cite{Erdos1959,Erdos1960,Erdos1961}. 
Using recursion equations, analytical results for 
the DSPL were obtained in different regimes, including sparse and dense 
networks of small as well as asymptotically large sizes.
The resulting distributions were found to 
be in good agreement with numerical simulations. 

ER networks are random graphs in which the degrees follow
a Poisson distribution and there are no degree-degree correlations
between connected pairs of nodes. 
In fact, ER networks 
can be considered as a maximum entropy ensemble under the constraint
that the mean degree is fixed.
Moreover, there is a much broader class of networks, 
named the configuration model, 
which generates maximum entropy ensembles
when the entire degree distribution is constrained
\cite{Newman2010,Newman2001,Fronczak2004,Molloy1995}.
The ER ensemble is equivalent to a configuration model in which
the degree distribution is constrained to be a Poisson distribution.
For any given degree distribution, one can produce an ensemble 
of configuration model networks and perform a statistical analysis
of its properties.
Therefore, the configuration model provides a general 
and highly powerful platform for the analysis of networks.
It is the ideal model to use as a null model
when one tries to analyze an empirical network of which the
degree distribution is known.
For a given empirical network, one constructs a configuration
model network of the same size and the same degree distribution.
Properties of interest 
such as the DSPL
\cite{Giot2003},
the betweenness centrality
\cite{Goh2003}
and the abundance of network motifs
\cite{Milo2002}
are compared between the two networks. 
The differences provide a rigorous test of the
systematic features of the empirical network vs. the random network.

A theoretical framework for the
study of the shell structure in configuration model networks
was developed in a series of papers
\cite{Kalisky2006,Shao2008,Shao2009}.
The shell
structure around the largest hub in a scale free network
was analyzed in Ref.
\cite{Kalisky2006}.
This approach was later extended into 
a general theory of the shell structure arond a random node
in a configuration model network
\cite{Shao2008,Shao2009}.
This formulation is based on recursion equations for the
number of nodes in each shell and for the degree distributions
in the shells. 
In the special case of the ER network, the results 
of Refs.
\cite{Shao2008,Shao2009} for
the number of nodes in each shell coincide with
those of Ref. 
\cite{Blondel2007}.

The shell structure around a random node in the configuration model
was recently utilized for the study of epidemic spreading 
\cite{Shao2015}.
In a study of biological networks,
the DSPL in a protein-protein interaction network
was analyzed and compared to a corresponding configuration model network
\cite{Giot2003}.
It was found that the distances in the configuration
model are shorter than in the original empirical network.
This highlights the features of the biological network which
tend to increase the distances.
These studies demonstrate the applicability of the configuration
model in the analysis of the structure and dynamics in empirical
networks.

In this paper we develop a theoretical framework,
based on the cavity approach 
\cite{Mezard1985,Mezard2003,Mezard2009,Ferraro2013},
for the calculation of the DSPL 
in networks which belong to the configuration model class. 
Using this framework we derive recursion equations for the calculation of
the DSPL in configuration model networks. 
We apply these equations to networks with 
degenerate, binomial and power-law degree 
distributions, and show that the results are in good agreement
with numerical simulations.
Using the tail-sum formula we calculate the mean and the variance
of the DSPL. Evaluating the discrete derivative of the tail 
distribution, we also obtain the mode of the DSPL.
These results provide closed form expressions for the central 
measures and dispersion measures of the DSPL in terms of the
moments of the degree distribution and the size of the network,
illuminating the connection between the two distributions.

The paper is organized as follows.
In Sec. II we present the class of configuration model networks.
In Sec. III we use the cavity approach to derive the recursion equations
for the calculation of the DSPL in these networks. 
In Sec. IV we consider properties of the DSPL such as 
the mean, mode and variance.
In Sec. V we present the results obtained from the
recursion equations 
for different network models
and compare them to numerical simulations. 
In Sec. VI we present a summary of the results.

\section{The configuration model}

The configuration model is a maximum entropy ensemble of
networks under the condition that the degree distribution
is imposed
\cite{Newman2001,Newman2010}.
Here we focus on the case of undirected networks, in 
which all the edges are bidirectional.
To construct such a network of $N$ nodes, one can draw
the degrees of all nodes from a desired degree
distribution 
$p(k)$,
$k=0,1,\dots,N-1$,
producing the degree sequence
$k_i$, $i=1,\dots,N$
(where $\sum k_i$ must be even).
The degree distribution 
$p(k)$
satisfies
$\sum_k p(k)=1$.
The mean degree over the ensemble of networks is
$c=\langle k \rangle=\sum_k k p(k)$,
while the average degree for a single instance of the network is
$\bar k  = \sum_i k_i/N$.
Here we consider networks which do not include isolated nodes,
namely $p(0)=0$.
This does not affect the applicability of the results, since 
the distribution of shortest path lengths is evaluated only
for pairs of nodes which reside on the same cluster, for which
the distance is finite.
Actually, if a network includes isolated nodes, one can discard them
by considering a renormalized degree distribution of the form
$p(k)/[1-p(0)]$, for $k=1,\dots,N-1$.

A convenient way to construct a configuration model network 
is to prepare the $N$ nodes such that each node, $i$, is 
connected to $k_i$ half edges
\cite{Newman2010}.
Pairs of half edges 
from different nodes
are then chosen randomly
and are connected to each other in order
to form the network. 
The result is a network with the desired degree sequence but
no correlations.
Note that towards the end of the construction
the process may get stuck.
This may happen in case that the only remaining pairs of half edges
are in the same node or in nodes which are already connected to each other.
In such cases one may perform some random reconnections 
in order to enable completion of the construction.

\section{Derivation of the recursion equations}

Consider a random pair of nodes, $i$ and $j$, in a 
network of $N$ nodes. Assuming that the two nodes
reside on the same connected cluster, they are likely to be
connected by a large number of paths. Here we focus on the shortest
among these paths (possibly more than one). 
More specifically, we derive recursion
equations for the length distribution
of these shortest paths.
To this end
we introduce the indicator function

\begin{equation}
\chi_N (d_{ij} > \ell) = 
\left\{
\begin{array}{ll}
1              & d_{ij}  >  \ell \\
0              & d_{ij} \le \ell,
\end{array}
\right.
\label{eq:chi}
\end{equation}

\noindent
where $d_{ij}$
is the length of the shortest path between nodes $i$ and $j$,
and $\ell$ is an integer. 
We also introduce the conditional indicator function

\begin{equation}
\chi_N (d_{ij} > \ell | d_{ij} > \ell - 1) = 
\frac{\chi_N(d_{ij} > \ell \cap d_{ij} > \ell-1)}
{\chi_N (d_{ij} > \ell-1)}. 
\label{eq:cond_chi}
\end{equation}

\noindent
Under the condition that the length $d_{ij}$ 
is larger than $\ell-1$,
this function indicates whether $d_{ij}$ is also larger than $\ell$.
If it is, the conditional indicator function $\chi=1$, otherwise 
(namely if $d_{ij} = \ell$) 
$\chi=0$.
In case the condition $d_{ij} > \ell-1$ is not satisfied,
the value of the conditional
indicator function is undetermined.
In order to extend this definition we adopt the convention
that in case the condition is not satisfied
the conditional indicator function takes the value  
$\chi_N (d_{ij} > \ell | d_{ij} > \ell - 1)=1$. 
We note that all the subsequent results are independent
of the value adopted here.
The indicator function
$\chi_N (d_{ij} > \ell)$
can be expressed as a product of the conditional indicator
functions in the form

\begin{equation}
\chi_N (d_{ij} > \ell) = 
\chi_N (d_{ij}>0)
\prod_{\ell^{\prime}=1}^{\ell}
\chi_N (d_{ij} > \ell^{\prime} | d_{ij} > \ell^{\prime}-1),
\label{eq:prod_chi}
\end{equation}

\noindent
where
$\chi_N (d_{ij}>0) = 1$,
since $i$ and $j$ are assumed to be two different nodes.

In the analysis below we calculate the mean of the
indicator function over an ensemble of networks to obtain the
distribution of shortest path lengths  
$P_N(d > \ell)$.
To this end 
we define the mean conditional indicator function
$m_i(\ell)\in [0,1]$,
obtained by averaging
the conditional indicator function 
$\chi_N (d_{ij} > \ell | d_{ij} > \ell - 1)$ 
over all suitable choices of
the final node, $j$: 

\begin{equation}
m_i(\ell) =
{\langle \chi_N (d_{ij}>\ell | d_{ij}>\ell-1) \rangle}_j.
\label{m-def2}
\end{equation}

\noindent
The averaging is done only over nodes $j$
which reside on the same cluster as node $i$ and for which
the condition
$d_{ij}>\ell-1$
is satisfied.

\begin{figure}
\begin{center}
\includegraphics[width=9cm]{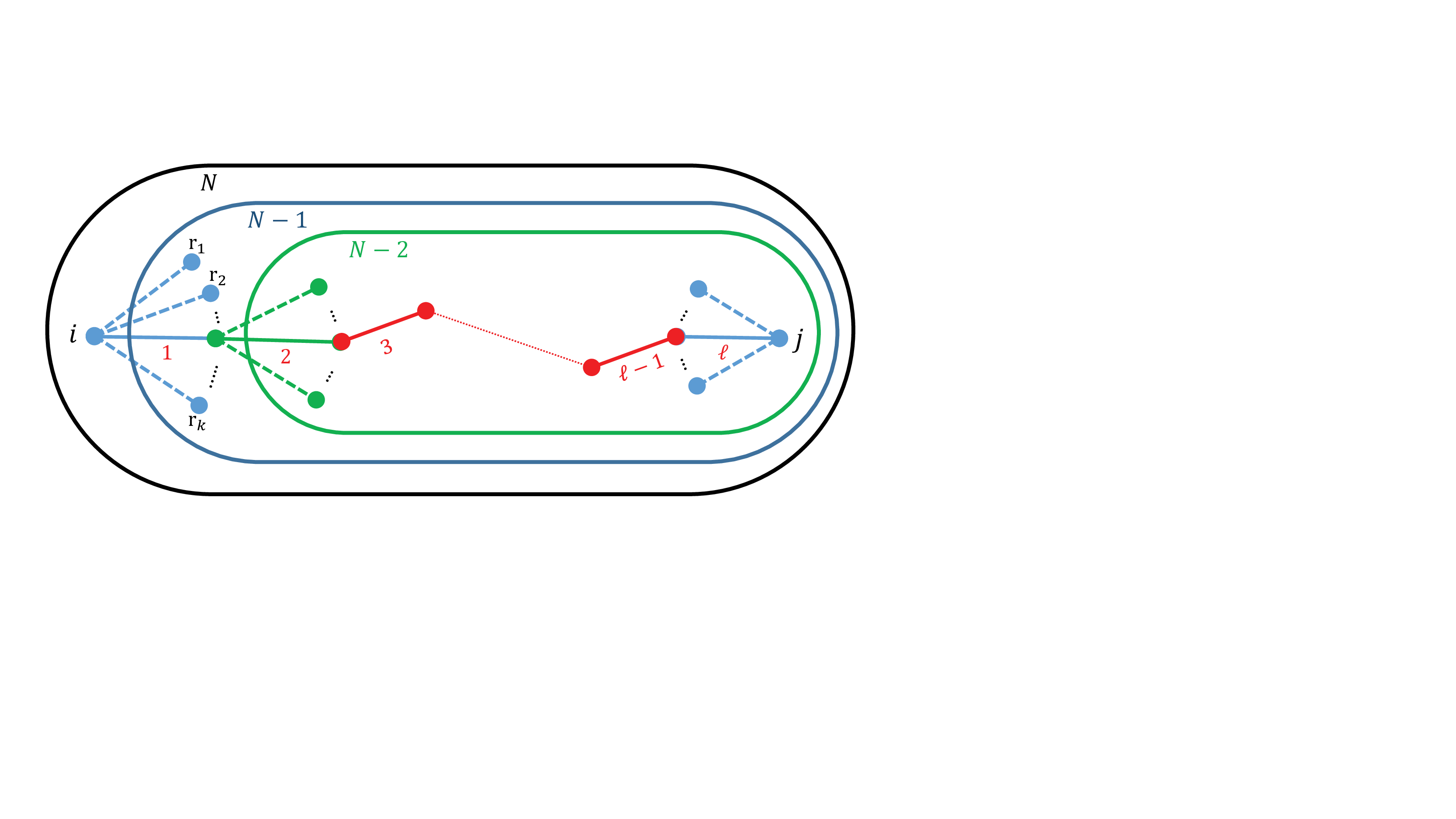}
\caption{
(Color online)
Illustration of the possible paths of 
length $\ell$ between two random nodes, $i$
and $j$, in a 
network of $N$ nodes. 
The first edge of such a path connects node $i$ 
to some other node, $r$, which may be any one 
of the $k$ neighbors of node $i$. 
The rest of the path, from node $r$ to node 
$j$ is of length $\ell-1$ and it 
resides on a smaller network
of $N-1$ nodes, from which node 
$i$ is excluded.
}
\label{fig:illustration}
\end{center}
\end{figure}

A path 
of length $\ell$
from node $i$ to node $j$ can be decomposed 
into a single edge connecting node $i$ and node 
$r \in \partial_i$
(where $\partial_i$ is the set of all nodes directly connected to $i$),
and a shorter path of length 
$\ell-1$ connecting $r$ and $j$.
Thus, the existence of a path of length $\ell$
between nodes $i$ and $j$
can be ruled out if there is no path of length
$\ell-1$ between any of the nodes 
$r \in \partial_i$,
and $j$
(Fig. 
\ref{fig:illustration}).
The conditional indicator functions for these paths 
of length $\ell-1$ are
$\chi_{N-1}^{(i)} (d_{rj}>\ell-1 | d_{rj}>\ell-2)$,
since they are embedded in a smaller 
network of $N-1$ nodes,
which does not include node $i$. 
The superscript $(i)$
stands for the fact that the node $r$ 
is reached by a link from node $i$.
This is often referred to as the cavity indicator function
\cite{Mezard1985,Mezard2003,Mezard2009,Ferraro2013}.
Similarly, we define the mean cavity indicator function as

\begin{equation}
m_r^{(i)}(\ell) =
{\langle \chi^{(i)}_N (d_{rj}>\ell | d_{rj}>\ell-1) \rangle}_j.
\label{m-def3}
\end{equation}

\noindent
This reasoning enables us to express the 
conditional indicator function
$\chi_N (d_{ij}>\ell | d_{ij}>\ell-1)$ 
as a product of conditional indicator 
functions for shorter paths between 
nodes
$r \in \partial_i$
and $j$

\begin{equation}
\chi_N (d_{ij}>\ell | d_{ij}>\ell-1) = 
\prod_{r \in \partial_i \texttt{\char`\\} \{j\}}
\chi_{N-1}^{(i)} (d_{rj}>\ell-1 | d_{rj}>\ell-2).
\label{cav-rec}
\end{equation}

\noindent
Under the assumption that the local structure of the
network is tree-like,
one can approximate the average of the product 
in Eq.
(\ref{cav-rec})
by the product of the averages.
This assumption is fulfilled in the limit of large networks.
In the analysis below we assume that $N \rightarrow \infty$ and
thus obtain recursion equations of the form

\begin{equation}
m_i(\ell)
= \prod_{r\in \partial_i \texttt{\char`\\} \{j\}} 
m_{r}^{(i)}(\ell-1). 
\label{eq:mi}
\end{equation}

\noindent
The mean cavity indicator function
$m_{r}^{(i)}(\ell)$
obeys a similar equation of the form

\begin{equation}
m_{r}^{(i)}(\ell)
= \prod_{s \in \partial_r \texttt{\char`\\} \{i,j\}} 
m_{s}^{(r)}(\ell-1). 
\label{eq:mir}
\end{equation}

\noindent
The number of neighbors
$r \in \partial_i$
is given by the degree, $k_i$, of node $i$,
while the number of neighbors 
$s \in \partial_r$
is given by the degree, $k_r$,
of node $r$.
Node $i$ is a randomly chosen node and thus
its degree, $k_i$, is drawn from $p(k)$.
Node $r$ is an intermediate node along the
path and its probability to be encountered
is proportional to its degree. Thus, its
degree, $k_r$, is drawn from the distribution
$(k/c) p(k)$, 
where $c$ takes care of the normalization.

Considering an ensemble of networks,
the variables
$m_i(\ell)$ 
and
$m_r^{(i)}(\ell)$,
which were defined for a specific node, $i$,
on a given instance of the network,
turn into the random variables
$m(\ell)$
and
$\tilde m(\ell)$, 
respectively.
These random variables 
are drawn from suitable probability distributions,
which respect the recursion equations
(\ref{eq:mi})
and
(\ref{eq:mir}).
We denote these
distributions by
$\pi_{\ell}(m) = Pr[m(\ell)=m]$
and
$\tilde \pi_{\ell}(m) = Pr[\tilde m(\ell)=m]$.
These distributions obey the equations

\begin{equation}
\pi_{\ell}(m) = \sum_{k=1}^{\infty} p(k)
\int_{0}^{1} \int_{0}^{1} \dots \int_{0}^{1} 
\prod_{\nu=1}^{k} \tilde \pi_{\ell-1}(m_{\nu}) {\rm d}m_{\nu} 
\delta \left(m-\prod_{\nu=1}^{k} m_{\nu} \right)
\label{eq:mdist}
\end{equation}

\noindent
and

\begin{equation}
\tilde \pi_{\ell}(m) = \sum_{k=1}^{\infty} \frac{k}{c} p(k)
\int_{0}^{1} \int_{0}^{1} \dots \int_{0}^{1} 
\prod_{\nu=1}^{k-1} \tilde \pi_{\ell-1}(m_{\nu}) {\rm d}m_{\nu} 
\delta \left(m-\prod_{\nu=1}^{k-1} m_{\nu} \right).
\label{eq:mdist2}
\end{equation}

\noindent
Eq.
(\ref{eq:mdist})
refers to the random node, $i$,
thus its degree is drawn from $p(k)$.
Eq. 
(\ref{eq:mdist2})
refers to intermediate nodes along the path,
thus the degrees are drawn from the distribution
$(k/c)p(k)$. 
An additional feature of the intermediate nodes is that
one of their edges is consumed by the incoming link, leaving only
$k-1$ links for the outgoing paths.

The expectation values of
$m(\ell)$
and
$\tilde m(\ell)$
over the graph ensemble yield 
the conditional probabilities 

\begin{equation}
m_{\ell} = P(d >\ell | d>\ell-1) = 
\int_{0}^{1} m \pi_{\ell}(m)  {\rm d}m
\label{eq:exp_m}
\end{equation}

\noindent
and

\begin{equation}
\tilde m_{\ell} =
\tilde P(d >\ell | d>\ell-1) = 
\int_{0}^{1} m \tilde \pi_{\ell}(m)  {\rm d}m.
\label{eq:exp_m2}
\end{equation}

\noindent
Plugging 
Eqs. (\ref{eq:mdist}) 
and
(\ref{eq:mdist2}) 
into
Eqs. (\ref{eq:exp_m})
and
(\ref{eq:exp_m2}),
respectively,
we obtain the recursion equations

\begin{equation}
m_{\ell} =
\sum_{k=1}^{\infty} p(k) 
(\tilde m_{\ell-1})^{k}
\label{eq:P_rec2}
\end{equation}

\noindent
and

\begin{equation}
\tilde m_{\ell} =
\sum_{k=1}^{\infty} \frac{k}{c} p(k) 
(\tilde m_{\ell-1})^{k-1},
\label{eq:P_rec2s}
\end{equation}

\noindent
which are valid for 
$\ell \ge 2$.
Recalling that $p(0)=0$,
Eqs.
(\ref{eq:P_rec2})
and
(\ref{eq:P_rec2s})
can be written
using the degree generating functions
\cite{Newman2001}

\begin{equation}
m_{\ell} = 
G_0 \left( 
\tilde m_{\ell-1}
\right)
\label{eq:P_recgen1}
\end{equation}

\noindent
and

\begin{equation}
\tilde m_{\ell} =
G_1 \left( 
\tilde m_{\ell-1}
\right),
\label{eq:P_recgen2}
\end{equation}

\noindent
where

\begin{equation}
G_0(x) = \sum_{k=0}^{\infty} p(k) x^k
\label{eq:genfunc}
\end{equation}

\noindent
and

\begin{equation}
G_1(x) = \sum_{k=0}^{\infty} \frac{k}{c} p(k) x^{k-1}.
\label{eq:cavgenfunc}
\end{equation}

\noindent
Eq. 
(\ref{eq:P_rec2})
can be understood intuitively as follows.
Consider the simplified scenario in which 
node $i$ is known to have a degree $k$.
In this case, excluding a path of length $\ell$
from $i$ to $j$ is equivalent to excluding a path
of length $\ell - 1$ from all $k$ neighbors of $i$
to $j$, namely
$m_{\ell} = (\tilde m_{\ell-1})^k$.
Such reasoning was applied in Ref.
\cite{Katzav2015},
to obtain the DSPL from a node with a given degree
to all other nodes in the network.
In practice, the degree of a random node is unknown,
and is distributed according to $p(k)$.
Therefore, Eq.
(\ref{eq:P_rec2})
averages over all possible degrees with suitable weights,
provided by $p(k)$.
Eq. 
(\ref{eq:P_rec2s})
can  be understood using a similar reasoning.

In the case of finite networks, we obtain

\begin{equation}
m_{N,\ell} =
\sum_{k=1}^{N-2} p(k) 
(\tilde m_{N-1,\ell-1})^k
\label{eq:P_finite}
\end{equation}

\noindent
and

\begin{equation}
\tilde m_{N,\ell} =
\sum_{k=1}^{N-2} \frac{k}{c} p(k) 
(\tilde m_{N-1,\ell-1})^{k-1},
\label{eq:P_finites}
\end{equation}

\noindent
for $\ell \ge 2$.
For $\ell=1$
we can directly obtain the results

\begin{equation}
m_{N,1} =
\sum_{k=1}^{N-1} p(k) 
\left( 1 - \frac{1}{N-1} \right)^k
\label{eq:P_rec5}
\end{equation}

\noindent
and

\begin{equation}
\tilde m_{N,1} =
\sum_{k=1}^{N-1} \frac{k}{c} p(k) 
\left(1 - \frac{1}{N-1} \right)^{k-1}.
\label{eq:P_rec5s}
\end{equation}

\noindent
The tail distribution of the shortest path lengths
can be expressed as a product of the form

\begin{equation}
P_N(d>\ell) =
P_N(d>0)
\prod_{\ell^{\prime}=1}^{\ell} 
P_N (d>\ell^{\prime} | d>\ell^{\prime}-1)
\equiv
P_N(d>0)
\prod_{\ell^{\prime}=1}^{\ell}
m_{N,\ell^{\prime}}.
\label{eq:prodcond}
\end{equation}

\noindent
Actually, since we choose two different nodes as the initial and final nodes, 
$P_N(d>0)=1$,
which further simplifies
Eq.
(\ref{eq:prodcond}).

\begin{figure}
\begin{center}
\includegraphics[width=13cm]{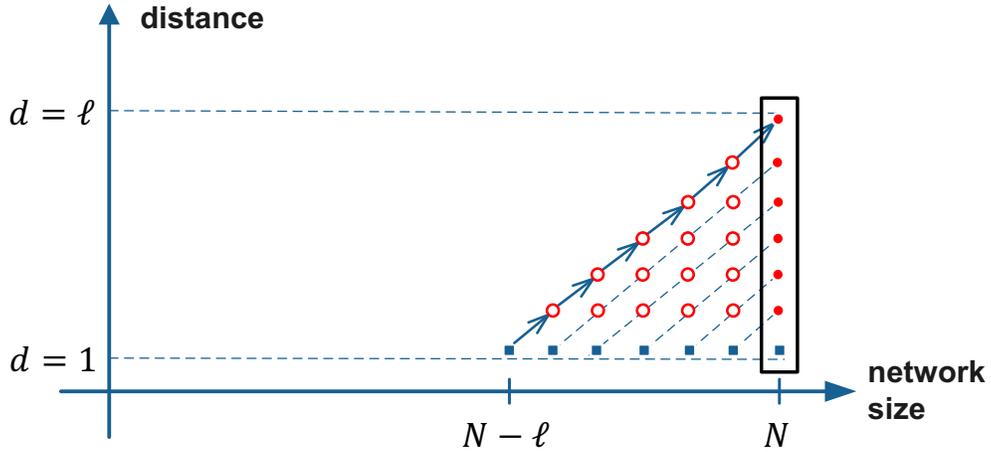}
\caption{
(Color online)
Illustration of the iteration process of the recursion 
equations 
(\ref{eq:P_finite}), 
and
(\ref{eq:P_finites}), 
which carry over along the diagonals
(empty circles). 
Starting from 
$\tilde m_{N-\ell^{\prime},1}$
(squares),
given by Eq.
(\ref{eq:P_rec5s}),
the iteration gives rise to
$m_{N,\ell^{\prime}}$
(full circles).
Eventually, $P_N(d>\ell)$
is obtained as a product of the
results in the right-most column
[Eq. (\ref{eq:prodcond})].
}
\label{fig:iterations}
\end{center}
\end{figure}

In Fig.  
\ref{fig:iterations}
we illustrate the way the recursion equations
are iterated $\ell^{\prime}-1$ times along the diagonal in order to obtain 
$m_{N,\ell^{\prime}}$.
Starting from 
$\tilde m_{N-\ell^{\prime},1}$
(squares),
Eq. (\ref{eq:P_finites})
is iterated $\ell^{\prime}-2$ times (empty circles),
followed by a single iteration 
(full circles)
of Eq.
(\ref{eq:P_finite}).
The desired value of
$P_N(d>\ell)$ is obtained
from Eq.
(\ref{eq:prodcond}).
This product runs from bottom to top
along the rightmost column
of Fig. 
\ref{fig:iterations}.

The probability distribution function,
namely,
the probability 
$P_N(\ell) = P_N(d=\ell)$ 
that the shortest path length between a random 
pair of nodes is equal to $\ell$ 
can be obtained from the tail distribution by

\begin{equation}
P_N(\ell)=P_N(d>\ell-1)-P_N(d>\ell),
\end{equation}
 
\noindent
for $\ell=1,2,\dots,N-1$.

It should be noted that 
Eqs.
(\ref{eq:mdist})
and
(\ref{eq:mdist2}),
presenting the distributions
$\pi_{\ell}(m)$
and
$\tilde \pi_{\ell}(m)$
enable the analysis of fluctuations of
the conditional probabilities within an
ensemble of networks with a given degree
distribution in the large $N$ limit.

\section{Properties of the DSPL}

The distribution of shortest path lengths,
$P_N(\ell)$,
can be characterized 
by its moments.
The $n$th moment, 
$\langle \ell^n \rangle$,
can be obtained 
using the tail-sum formula
\cite{Pitman1993}

\begin{equation}
\langle \ell^n \rangle = 
\sum_{\ell=0}^{N-2} [(\ell+1)^n - \ell^n] P_N(d>\ell).
\label{eq:tail_sum}
\end{equation}

\noindent
Note that the sum in 
Eq.
(\ref{eq:tail_sum})
does not extend to $\infty$ because the longest possible shortest path in a network of size $N$ is $N-1$.
The average distance between pairs of nodes
in the network is given by the first moment

\begin{equation}
\langle \ell \rangle = \sum_{\ell=0}^{N-2} P_N(d>\ell).
\label{eq:mean_ell}
\end{equation}

\noindent
The average distance between nodes in configuration model
networks has been studied extensively
\cite{Newman2001,Chung2002,Chung2003,Hofstad2005,Esker2006,Bollobas2007,Hofstad2007,Esker2008}.
It was found that 

\begin{equation}
\langle \ell \rangle \simeq 
\frac{\ln N}{\ln \left(\frac{\langle k^2 \rangle - \langle k \rangle}
{\langle k \rangle}\right)} + {\cal O}(1). 
\label{eq:mean_ellold}
\end{equation}

\noindent
The width of the distribution 
can be characterized by the
variance
$\sigma_{\ell}^2 = \langle \ell^2 \rangle - \langle \ell \rangle^2$,
where
 
\begin{equation}
\langle \ell^2 \rangle = \sum_{\ell=0}^{N-2} (2 \ell + 1) P_N(d>\ell).
\label{eq:secmom_ell}
\end{equation}

\noindent
In addition to the average distance
$\langle \ell \rangle$,
another common measure of the typical distance
between nodes in the network is the mode.
Here we present a way to extract 
the mode of 
$P_N(\ell)$
directly from the 
recursion equations, 
in the limit of a large network. 
It is based on the following observations:
(a) The tail-distribution, 
$P_N(d>\ell)$,
is a sigmoid function, i.e. it 
starts at $1$ at the origin and drops to $0$ at infinity. 
The transition between the two levels occurs over a 
relatively narrow interval;
(b) Actually,
$P_N(d>\ell)$
can be expressed as a product of 
conditional probabilities of the form
$m_{N,\ell^{\prime}}$,
where each term has the form of a sigmoid function 
[Eq. (\ref{eq:prodcond})]. 
Therefore, the product becomes an even 
sharper sigmoid function, and to a good approximation its maximal slope 
is determined by the the last term in the product.
Therefore, in the analysis below we focus on the 
conditional probability
$m_{N,\ell}$.

Considering the large $N$ limit
we can use the recursion equations
(\ref{eq:P_recgen1})
and
(\ref{eq:P_recgen2}).
The generating functions satisfy
$G_0(1)=G_1(1)=1$, thus
both equations exhibit a (repelling) fixed point at
$m_{\ell} = \tilde m_{\ell} = 1$.
Note that in this formulation,
the network size $N$ 
does not appear explicitly in the recursion equations, 
but only enters through the initial conditions,
given by Eqs.
(\ref{eq:P_rec5})
and
(\ref{eq:P_rec5s}).
For simplicity, we approximate Eqs.
(\ref{eq:P_rec5})
and
(\ref{eq:P_rec5s})
by

\begin{equation}
m_{1} \simeq 1 - \frac{c}{N-1} + {\cal O} \left(\frac{1}{N^2}  \right),
\end{equation}

\noindent
and

\begin{equation}
\tilde m_{1} \simeq 1 - 
\frac{\langle k^2 \rangle - \langle k \rangle}{\langle k \rangle (N-1)}
+ {\cal O} \left(\frac{1}{N^2}  \right),
\end{equation}

\noindent
respectively.
For networks which are not too dense,
these values are only slightly smaller than $1$.
Therefore, the linearized versions of Eqs. 
(\ref{eq:P_recgen1})
and
(\ref{eq:P_recgen2})
hold as long as 
$m_{\ell}$ 
and
$\tilde m_{\ell}$
are
sufficiently close to $1$.
Note that these expressions 
require that the second moment 
$\langle k^2 \rangle$ 
would be finite.
This condition may limit the validity of the derivation presented below
to networks for which 
$\langle k^2 \rangle$ is bounded.
Thus, networks for which
$\langle k^2 \rangle$ diverges require special attention.

The location of the
maximum value of the probability distribution function
(namely the mode) is obtained at the point where the 
tail distribution falls most sharply. Up to that point 
the linear approximation holds quite well.
This motivates us to perform the analysis in terms
of the deviations

\begin{equation}
\epsilon_{\ell} = 1 - m_{\ell},
\label{eq:eps}
\end{equation}

\noindent
and

\begin{equation}
\tilde \epsilon_{\ell} = 1 - \tilde m_{\ell}.
\label{eq:epst}
\end{equation}

\noindent
Linearizing Eqs.
(\ref{eq:P_recgen1})
and 
(\ref{eq:P_recgen2})
in terms of 
$\epsilon_{\ell}$
and
$\tilde \epsilon_{\ell}$,
respectively,
we obtain 

\begin{equation}
\epsilon_{\ell} = 
\langle k \rangle \tilde \epsilon_{\ell-1},
\label{eq:epsilonl}
\end{equation}

\noindent
and

\begin{equation}
\tilde \epsilon_{\ell} = 
\left[\frac{\langle k^2 \rangle - \langle k \rangle}{\langle k \rangle}\right]^{\ell-1}
\tilde \epsilon_1,
\label{eq:epsilonls}
\end{equation}

\noindent
for any  
$\ell \ge 2$,
where
$\tilde \epsilon_1 = 
{(\langle k^2 \rangle - \langle k \rangle)}/{[\langle k \rangle (N-1)]}$.
Our aim is to determine the value of $\ell$ at which
the reduction in  
$m_{\ell}$
is maximal.
We denote the discrete derivative

\begin{equation}
\Delta P = 
m_{\ell-1} - m_{\ell}.
\label{eq:Deltap}
\end{equation}

\noindent
Using the recursion equations
(\ref{eq:P_recgen1})
and
(\ref{eq:P_recgen2}),
we can express this as

\begin{equation}
\Delta P = 
G_0(\tilde m_{\ell-2}) - G_0[G_1(\tilde m_{\ell-2})],
\label{eq:Deltap3}
\end{equation}

\noindent
and we are therefore interested in the value of $x$,
denoted by $x_{max}$,
at which the function
$\Delta P(x) = G_0(x) - G_0[G_1(x)]$
is maximal. 
This is determined by the solution of the extremum condition

\begin{equation}
\frac{d\Delta P}{dx} = 
G_0^{\prime}(x) - G_0^{\prime}[G_1(x)]G_1^{\prime}(x) = 0.
\label{eq:Deltap4}
\end{equation}

\noindent
As long as $x_{max}$ is close to $1$ 
we can use the linear approximation leading to Eq.
(\ref{eq:epsilonls}),
in which case we can equate 
$\tilde \epsilon_{\ell_{mode}-2+{\cal O}(1)} = 1 - x_{max}$, 
where the ${\cal O}(1)$ term comes from the fact that we are using
a linearized equation while potentially higher order 
corrections should have been considered. This term is small and 
could be omitted when $x_{max}$ is close to $1$, which is 
the situation in various known cases.
Combining this result with Eq. 
(\ref{eq:epsilonls})
we obtain

\begin{equation}
\ell_{mode} = 
\frac{\ln \left[(N-1)(1- x_{max})\right]}
{\ln \left( \frac{\langle k^2 \rangle - 
\langle k \rangle}{\langle k \rangle}\right)} + 2 +{\cal O}(1).
\label{eq:mode}
\end{equation}

\noindent
It is interesting to note that the mode exhibits the same
scaling with the network size as the average distance shown
in Eq.
(\ref{eq:mean_ellold}).
This analysis is in the spirit of the 
renormalization group approach,
where the flow of an initial small deviation 
from the critical temperature 
(here from the fixed point $m=1$), 
under the linearized renormalization transformation 
determines the scaling behaviour of the system.

\section{Analysis of Network Models}

To examine the recursion equations we apply them to the 
calculation of the DSPL in 
configuration model networks with different choices of the
degree distribution.
The results are compared to numerical simulations. 
In these simulations we generate instances of the configuration
model networks with the required degree distribution.
We then calculate the distances between all pairs of nodes in each
network and generate a histogram. 
The process is repeated over a large number network instances.
In case that the network includes more than one connected 
cluster we take into account only the distances between
pairs of nodes which reside on the same cluster.
The DSPL obtained from the numerical simulations is normalized
accordingly.

To cover a broad class of networks, we consider configuration models
which exhibit narrow as well as broad degree distributions.
For networks with narrow degree distributions we study the 
the regular network
(degenerate distribution) and networks with 
a binomial distribution.
For networks with broad degree distributions we study configuration
models with power-law degree distributions (scale-free networks).
A detailed analysis of the 
distributions of shortest path lengths in these
configuration models is presented below.

\subsection{Regular Networks}

The simplest case of the configuration model is the regular graph,
in which the degree distribution is
$p(k)=\delta_{k,c}$, namely
all $N$ nodes have the same degree, 
(where $c \ge 2$ and $N c$ is even). 
For $c=2$ the network consists
only of loops, while for $c \ge 3$ more complex network
structures appear.
The random regular graph ensemble has been studied extensively
and enjoys many analytical results
\cite{Wormald1999}. 
In particular, there is an interesting phase transition at $c=3$
above which the network becomes connected with probability $1$
in the asymptotic limit.

In case of the regular graph
the recursion equations 
(\ref{eq:P_finite})
and
(\ref{eq:P_rec5})
take the form

\begin{equation}
m_{N,\ell} =
(\tilde m_{N-1,\ell-1})^c
\label{eq:P_regularnc}
\end{equation}

\noindent
and

\begin{equation}
m_{N,1} =
\left( 1 - \frac{1}{N-1} \right)^c,
\label{eq:P_regularlast}
\end{equation}

\noindent
respectively.
The subsequent equations,
derived from Eqs.
(\ref{eq:P_finites})
and
(\ref{eq:P_rec5s})
take the form

\begin{equation}
\tilde m_{N,\ell} =
(\tilde m_{N-1,\ell-1})^{c-1}
\end{equation}

\noindent
and

\begin{equation}
\tilde m_{N,1} =
\left(1 - \frac{1}{N-1} \right)^{c-1}.
\end{equation}

\noindent
The iteration of these equations gives rise to a closed form
equation for the conditional probabilities

\begin{equation}
P_N\big(d>\ell | d>\ell-1) = 
m_{N,\ell} =
\left(1-\frac{1}{N-\ell} \right)^{c(c-1)^{({\ell-1})}}.
\label{eq:reg}
\end{equation}

\noindent
Inserting the conditional probabilities into Eq.
(\ref{eq:prodcond}),
and using the approximation 
$N - \ell \simeq N$,
we obtain the tail distribution

\begin{equation}
P_N\big(d>\ell) = 
\exp \left[{- \frac{c (c-1)^{\ell}}{N(c-2)}}\right],
\label{eq:regulartail}
\end{equation}

\noindent
in agreement with Eq. (1.10) in Ref.
\cite{Hofstad2005}.

Actually, in this case, 
Eqs.
(\ref{eq:mdist})
and
(\ref{eq:mdist2}),
describing the 
fluctuations in the ensemble 
in the large $N$ limit,
can be solved analytically 
yielding

\begin{equation}
\pi_{\ell}(m) = \delta 
\left[m-\left(1-\frac{1}{N} \right)^{c(c-1)^{({\ell-1})}} \right].
\label{eq:reg_delta}
\end{equation}

\noindent
This means that in regular networks, for sufficiently large $N$,
the fluctuations are negligible.

The mean distance, 
$\langle \ell \rangle$,
for the regular graph thus
takes the form

\begin{equation}
\langle \ell \rangle = 
\sum_{\ell=0}^{N-2}
e^{- \frac{c(c-1)^{\ell}}{N(c-2)}}. 
\label{eq:regularmean}
\end{equation}

\noindent
It is useful to define

\begin{equation}
s = \left\lfloor \frac{\ln N}{\ln (c-1)} \right\rfloor,
\label{eq:s}
\end{equation}

\noindent
where $\lfloor x \rfloor$
is the integer part of $x$.
It is easy to see that for 
$\ell = 0,1,\dots,s$,
the exponents on the right hand side 
of Eq.
(\ref{eq:regularmean})
are very close to $1$,
while for 
$\ell > s$
these exponents are quickly reduced.
Therefore, to a very good approximation
$\langle \ell \rangle = \ln N / \ln (c-1)$.
In order to obtain a more systematic approximation of
$\langle \ell \rangle$
we take into account explicitly a few terms around
$\ell = s$
in Eq.
(\ref{eq:regularmean}).
For example, taking three terms explicitly we obtain

\begin{equation}
\langle \ell \rangle = 
(s-1) + 
\sum_{\ell=s-1}^{s+1}
e^{- \frac{c(c-1)^{\ell}}{N(c-2)}}. 
\label{eq:lav3}
\end{equation}

\noindent
One can easily improve the approximation by including additional 
explicit terms to the right and left of 
$\ell = s$.
Higher order moments can be evaluated in a similar fashion,
yielding

\begin{equation}
\langle \ell^n \rangle = 
(s-r)^n + 
\sum_{\ell=s-r}^{s+r}
[(\ell+1)^n - \ell^n]
e^{- \frac{c(c-1)^{\ell}}{N(c-2)}}, 
\label{eq:lavn}
\end{equation}

\noindent
where $r$ is the number of terms taken into account 
explicitly on the right and on the left.
The variance of $P_N(\ell)$ is thus

\begin{equation}
\sigma_{\ell}^2 = 
\sum_{\ell^{\prime}=-r}^{r}
(2 \ell^{\prime}+ 2 r + 1)
e^{- \frac{c(c-1)^{s+\ell^{\prime}}}{N(c-2)}}
-
\left[\sum_{\ell^{\prime}=-r}^{r}
e^{- \frac{c(c-1)^{s+\ell^{\prime}}}{N(c-2)}}\right]^2.
\label{eq:lsigman}
\end{equation}

\begin{figure}[H]
\begin{center}
\includegraphics[width=0.7\columnwidth]{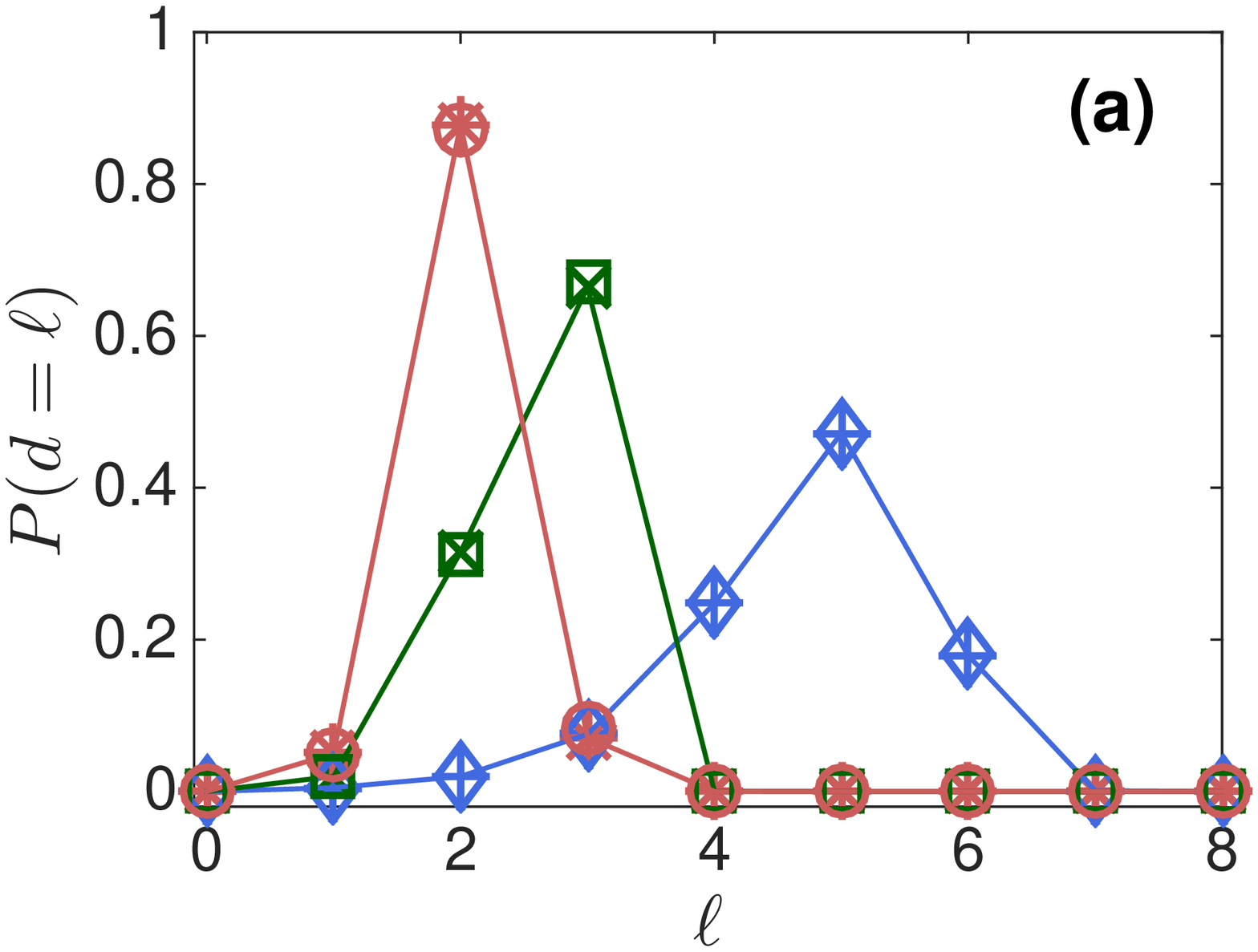}
\includegraphics[width=0.7\columnwidth]{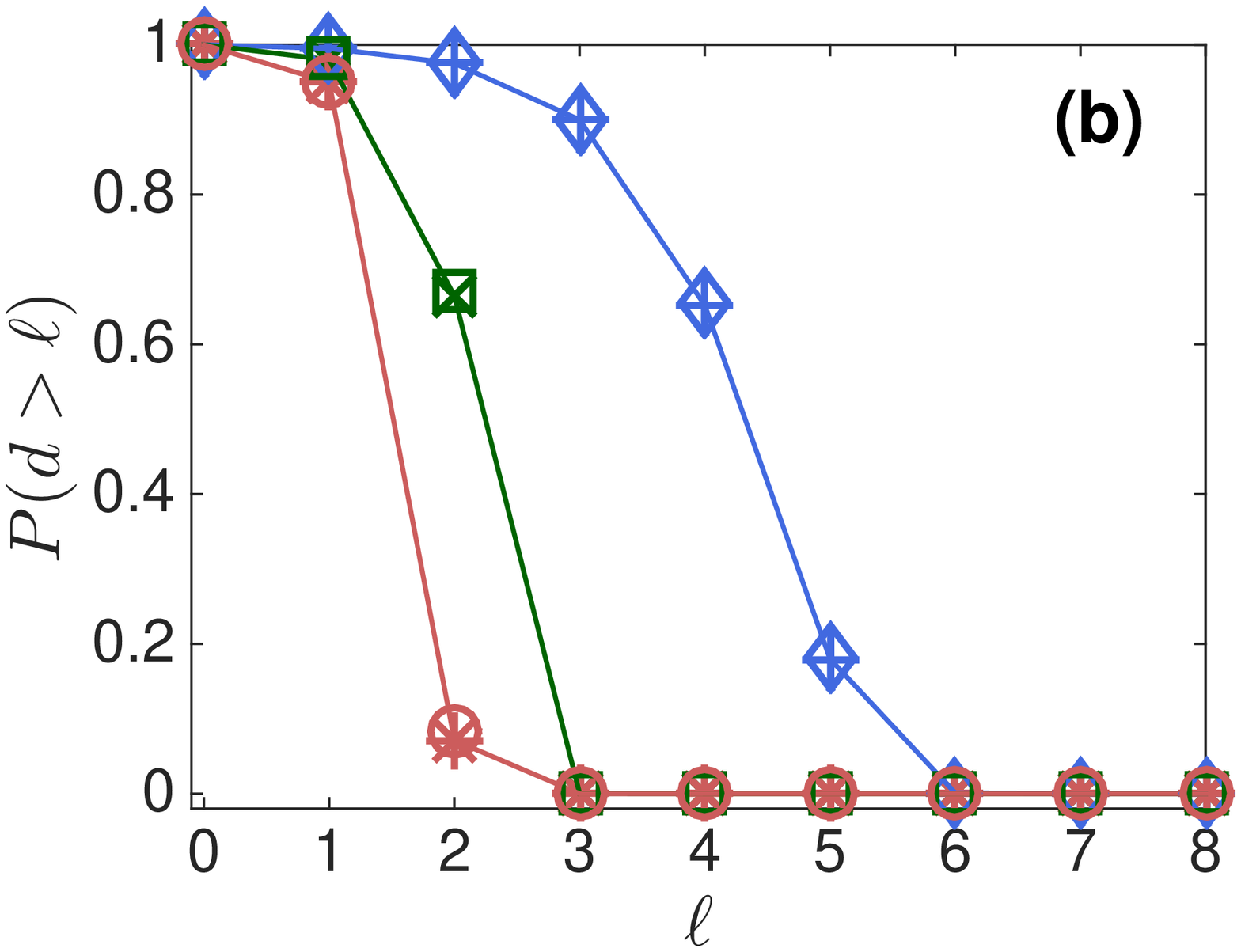}
\caption{
(Color online)
Distribution of shortest path lengths in a 
regular graph.
The results of the recursion equations for 
$P(\ell)$ (a) and 
$P(d>\ell)$ 
(b), 
for 
$c=5$, $20$ and $50$ 
($\Diamond$, $\square$ and 
$\bigcirc$ , respectively), 
fit well the 
numerical results 
(+, $\times$ and $\ast$, respectively).
The numerical results were averaged over $50$ 
graph instances in a graph of size 
$N=1000$.
}
\label{fig:regular}
\end{center}
\end{figure}

\noindent
In Fig. 
\ref{fig:regular}
we present the DSPL for regular networks
of $N=1000$ nodes, with $c=5$, $20$ and $50$,
obtained from Eq.
(\ref{eq:regulartail}).
The probability distribution function $P(d=\ell)$
is shown in 
Fig. 
\ref{fig:regular}(a)
and the tail distribution $P(d>\ell)$
is shown in 
Fig. \ref{fig:regular}(b).
The results are compared with computer simulations
showing excellent agreement.

\begin{figure}[H]
\begin{center}
\includegraphics[width=0.7\columnwidth]{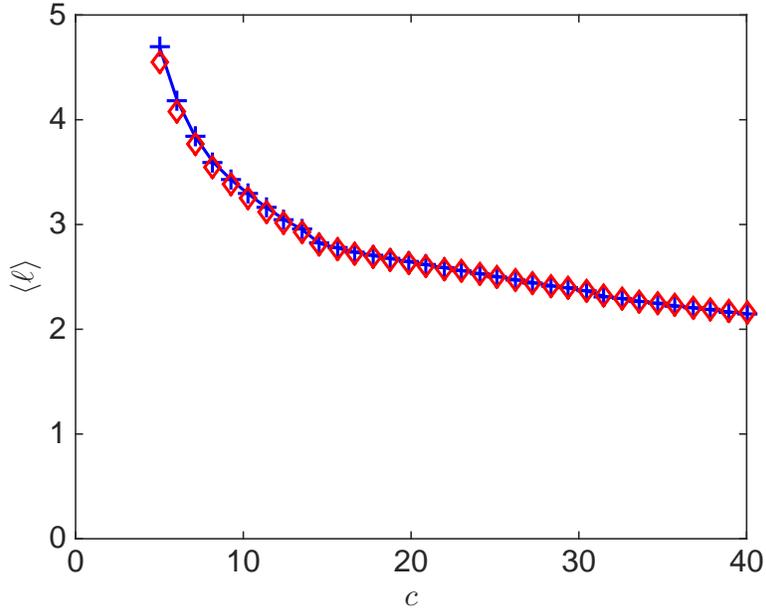}
\caption{
(Color online)
Mean shortest path length,
$\langle \ell \rangle$,
vs. the degree, $c$,
in a regular graph
of size $N=1000$. 
The results of the recursion equations 
($\Diamond$) 
are in very good agreement with the
numerical results (+). 
The numerical results were averaged over $50$ 
graph instances.
}
\label{fig:ell_c}
\end{center}
\end{figure}

In 
Fig. \ref{fig:ell_c}
we present the mean distance in regular graphs of
$N=1000$ nodes vs. the degree $c$,
obtained from the recursion equations
($\diamond$).
The results are in excellent agreement with numerical 
simulations ($+$).
As expected, the average distance decreases
logarithmically as $c$ is increased, in very good
agreement with the exact result 
$\langle \ell \rangle = \ln N / \ln (c-1)$.

For the regular graph,
$\langle k \rangle =c$
and
$\langle k^2 \rangle =c^2$.
Plugging the degenerate degree distribution
$p(k) = \delta_{k,c}$
into Eqs.
(\ref{eq:genfunc})
and
(\ref{eq:cavgenfunc})
we obtain that for the regular network
$G_0(x)=x^c$
and
$G_1(x)=x^{c-1}$.
Since the distribution
$P_N(\ell)$ 
for the regular network
is narrow, 
one expects the mode 
$\ell_{\rm mode}$
of this distribution
to follow closely the mean value
$\langle \ell \rangle$
and to increase logarithmically as a function of $N$. 
Here we evaluate 
$\ell_{\rm mode}$
using Eq.
(\ref{eq:mode}).
Inserting
$x_{max} = (c-1)^{-1/(c-1)}$
into Eq. 
(\ref{eq:mode})
we obtain 

\begin{equation}
\ell_{mode} = \frac{\ln N}{\ln (c-1)} +{\cal O}(1).
\label{eq:lmoder}
\end{equation}

\begin{figure}[H]
\begin{center}
\includegraphics[width=0.6\columnwidth]{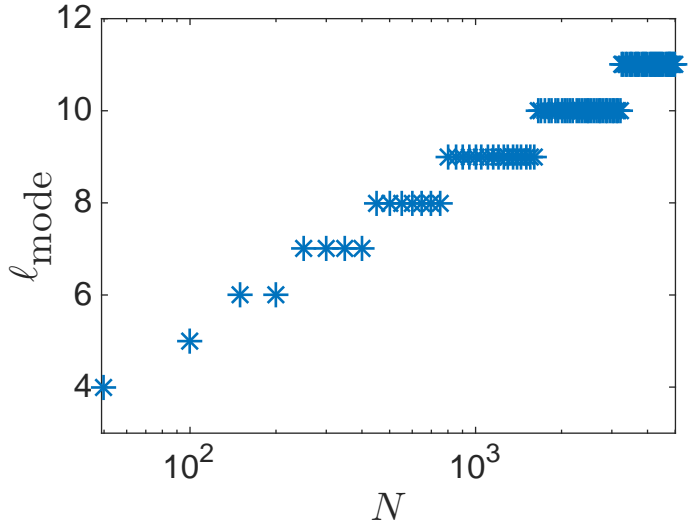}
\caption{
(Color online)
The mode of the distribution 
$P_N(\ell)$
as a function of the network size, $N$, 
for a regular network of degree $c=3$.
Overall, 
the mode 
scales logarithmically with the network size.
However, on a finer scale it forms steps due to 
the discreteness of the distance $\ell$.
}
\label{fig:mode}
\end{center}
\end{figure}

\noindent
Unlike 
$\langle \ell \rangle$
the mode takes only integer values.
Therefore, it must take the form of a step function vs. $N$.
In Fig.  
\ref{fig:mode}
we present
$\ell_{\rm max}$
vs. $N$ on a semi-logarithmic scale.
The general trend indeed satisfies
$\ell_{\rm max} \sim \ln N$,
but the graph is decorated by steps at integer
values of
$\ell_{\rm max}$.

\subsection{Networks with Binomial Degree Distributions}

To further examine the recursion equations,
we extend the analysis to networks which exhibit 
a narrow or bounded degree distribution,
with an average
$\langle k \rangle = c$
and variance
$\sigma_k^2$.
Since the degree distribution, 
$p(k)$, is a discrete distribution,
the binomial distribution 

\begin{equation}
p(k) = {n \choose k}p^k (1-p)^{n-k}, 
\label{eq:binomial}
\end{equation}

\noindent
where $n$ is an integer and $0 < p < 1$,
is particularly convenient.
Its mean is given by
$\langle k \rangle = np$
and its variance is given by
$\sigma_k^2 = np(1-p)$.
In order to obtain desired values of
$\langle k \rangle$ 
and 
$\sigma_k^2$,
we choose the
parameters 
$n$ and $p$ 
according to

\begin{equation}
n = {\rm Round} \left( \frac{\langle k \rangle^2}{\langle k \rangle - \sigma_k^2} \right),
\end{equation}

\noindent
where ${\rm Round}(x)$ is the nearest integer to $x$,
and

\begin{equation}
p = \frac{\langle k \rangle - \sigma_k^2}{\langle k \rangle }.
\end{equation}

\begin{figure}[H]
\includegraphics[width=0.7\columnwidth]{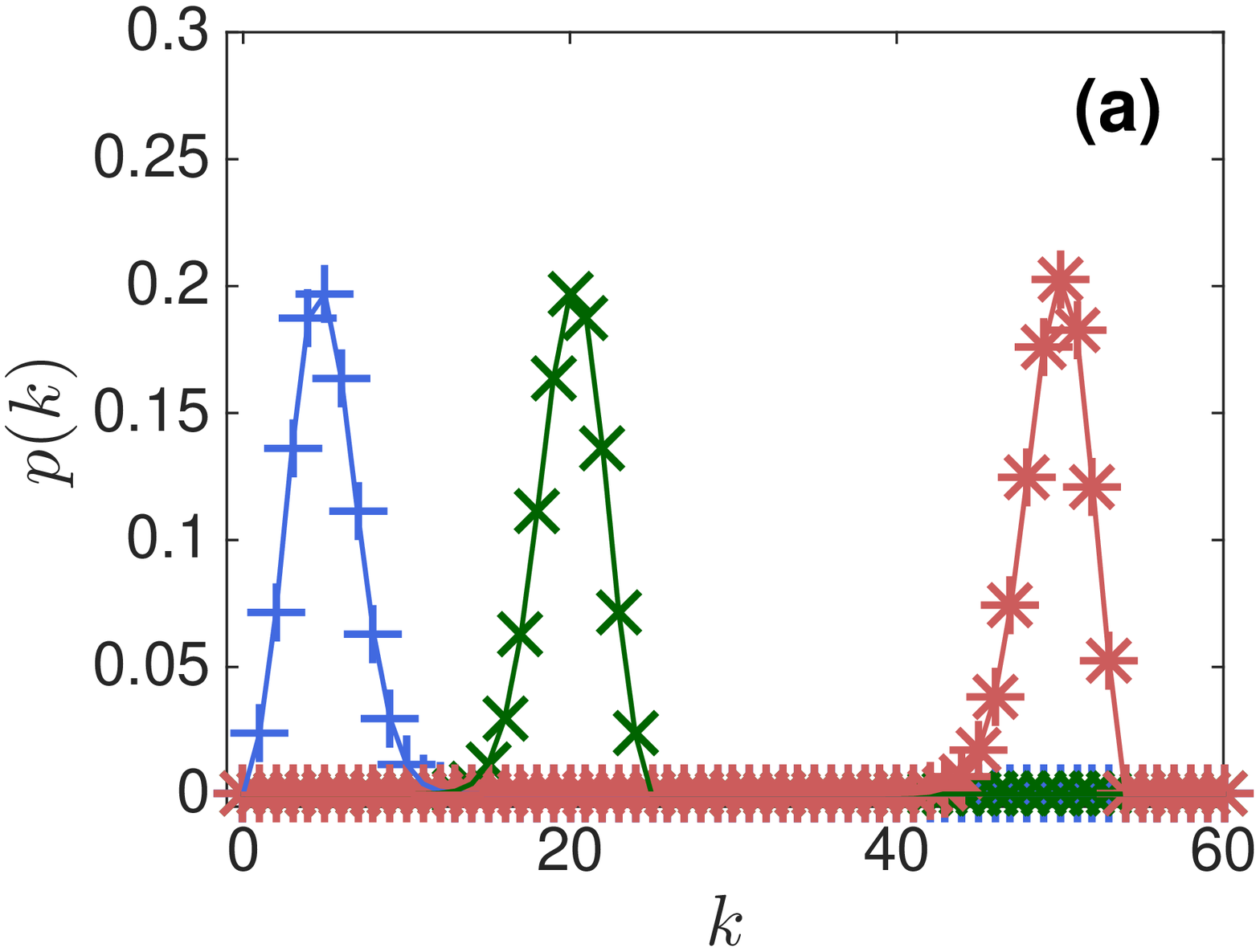}
\includegraphics[width=0.7\columnwidth]{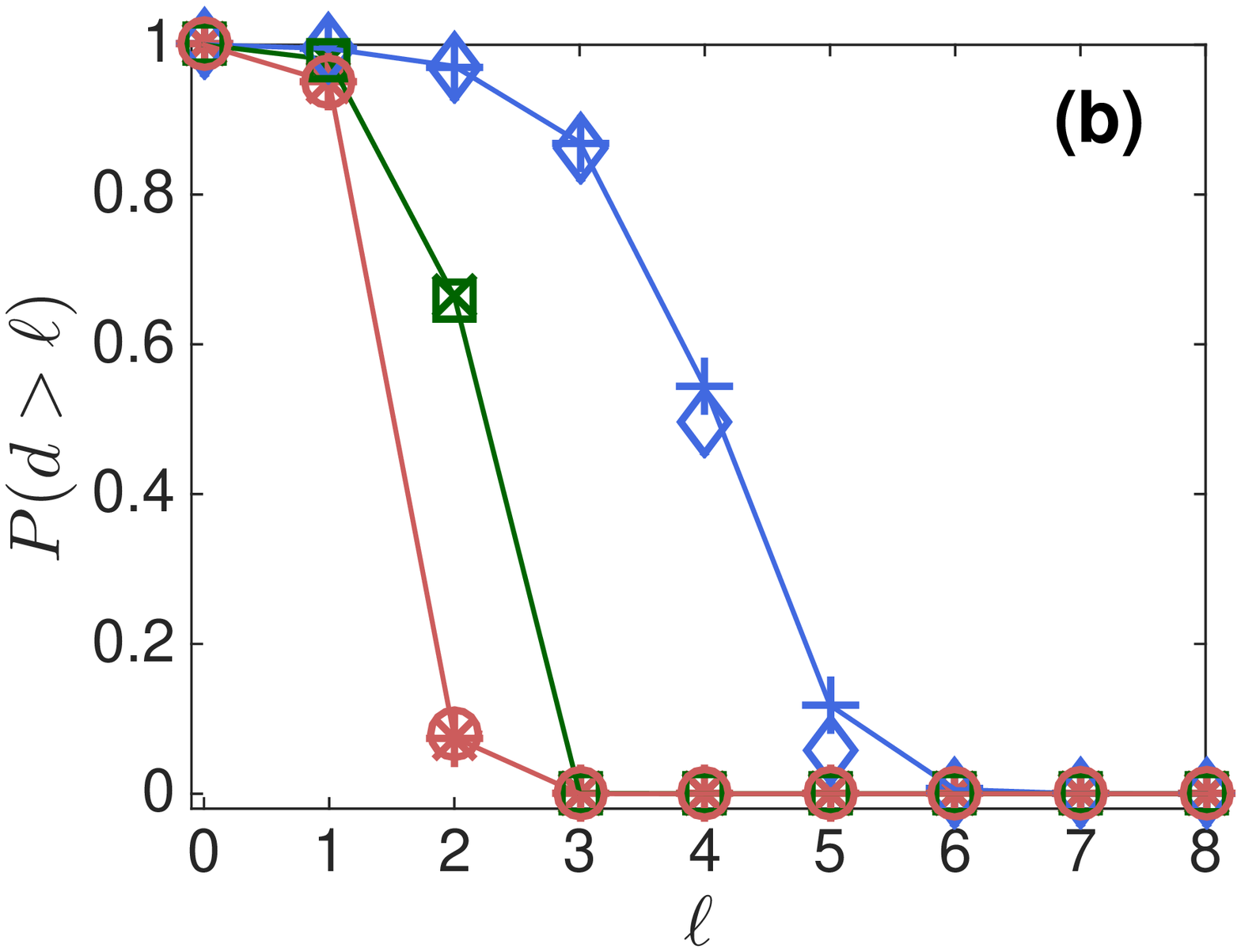}
\caption{
(Color online)
(a) The degree distributions of three networks of size $N=1000$,
where $p(k)$ was drawn from binomial distributions with means
$c=5$ , $20$ and  $50$ 
(+, $\times$ and $\ast$, 
respectively),
for which the standard deviation is 
$\sigma_k=4$.
The results were obtained from numerical simulations,
averaging over $50$ graph instances.
These results verify the construction of the configuration model network. 
(b) The tail distribution
$P(d>\ell)$,
obtained from the recursion equations
($\Diamond$, $\square$ and $\bigcirc$, 
respectively), 
and from numerical simulations
(+, $\times$ and $\ast$ , 
respectively),
for the three networks described above.
It is observed that as the mean degree is increased,
the average distance decreases.
}
\label{fig:binom}
\end{figure}

\noindent
It is important to note that the parameter, $n$, is not related
to the network size, $N$, and can be either larger or smaller than $N$.
However, one should choose a combination of 
$n$ and $p$ for which the
probability, $p(k)$, 
for $k > N-1$ is vanishingly small,
otherwise a truncation will be needed,
which will deform the distribution.
In Fig. 
\ref{fig:binom}(a)
we present the binomial degree distributions of three 
ensembles of networks of $N=1000$ nodes, 
$c=5$ (+), $20$ ($\times$) and $50$ ($\ast$)
and $\sigma_k = 4$.
In Fig. 
\ref{fig:binom}(b)
we present the tail distributions
$P(d>\ell)$
for these three network ensembles,
obtained from the recursion equations for
$c=5$ ($\diamond$), $20$ ($\square$) and $50$ ($\circ$).
The results are found to be in very good agreement with
numerical simulations,
(+, $\times$ and $\ast$, 
respectively),
except for the case of $c=5$, 
where some small deviations are observed.
These deviations are due to the fact that
in sparse networks the weight
of the small, isolated clusters may be non-negligible even above the
percolation threshold. 
This gives rise to some discrepancy 
between the theoretical and the numerical results for
$P(d > \ell)$ 
for small values of $c$.

Plugging the binomial degree distribution
of Eq. 
(\ref{eq:binomial})
into Eqs.
(\ref{eq:genfunc})
and
(\ref{eq:cavgenfunc})
we obtain that
$G_0(x) = [1-p(1-x)]^n$
and
$G_1(x) = [1-p(1-x)]^{n-1}$.
In the asymptotic limit,
where $n \gg 1$, this expression
converges to
$G_0(x) \simeq G_1(x) \simeq e^{-c(1-x)}$.

Here we evaluate 
$\ell_{\rm mode}$
for a network with a binomial degree distribution
using Eq.
(\ref{eq:mode}).
For such networks
$x_{max}=1-\ln c/c$.
Inserting the results above into Eq.
(\ref{eq:mode})
we obtain 

\begin{equation}
\ell_{mode} = \frac{\ln N}{\ln c} +{\cal O}(1).
\label{eq:moden}
\end{equation}

\noindent
Note that Eqs. 
(\ref{eq:lmoder}) 
and 
(\ref{eq:moden}) 
differ in their denominators,
where the former is 
$\ln (c-1)$
while the latter is
$\ln c$.
The reason for this difference comes from the fact 
that in the regular network each node has exactly 
$c$ neighbours, and so only $c-1$ of them actually 
connect inner to outer shells. However, in the 
binomial case (as in the ER case), each neighbour
of the initial node has on average an extra edge, 
and thus $c$ edges connect an inner shell to an outer shell.

\subsection{Networks with Power-Law Degree Distributions}

Studies of empirical networks revealed that many of them exhibit
power-law degree distributions of the form 
$p(k) \sim k^{-\gamma}$,
where 
$2 < \gamma < 3$.
This is the range of values of 
$\gamma$
for which the average degree is bounded 
but its variance diverges in the infinite system limit. 
To construct a configuration model 
network with a power-law distribution 
$p(k)$, we first choose a lower cutoff 
$k_{min} \ge 1$
and an upper cutoff 
$k_{max} \le N-1$.
We then draw the degree sequence 
$k_i$, $i=1,\dots,N$
from the distribution

\begin{equation}
p(k) = A k^{-\gamma}, 
\label{eq:power_law}
\end{equation}

\noindent
where the normalization coefficient is

\begin{equation}
A = [\zeta(\gamma,k_{min})-\zeta(\gamma,k_{max}+1)]^{-1}, 
\label{eq:power_lawcoeff}
\end{equation}

\noindent
and
$\zeta(s,a)$ is
the Hurwitz zeta function
\cite{Olver2010}. 
In the analytical calculations we insert 
$p(k)$ from
Eq.
(\ref{eq:power_law})
into the recursion equations in order to obtain the distribution
of shortest path lengths for the ensemble of networks produced using
this degree distribution.
In the numerical simulation we repeatedly draw degree sequences
from this distribution,
produce instances of configuration model networks, calculate the
distribution of shortest path lengths in these networks and average
over a large number of instances.

\begin{figure}[H]
\begin{center}
\includegraphics[width=0.7\columnwidth]{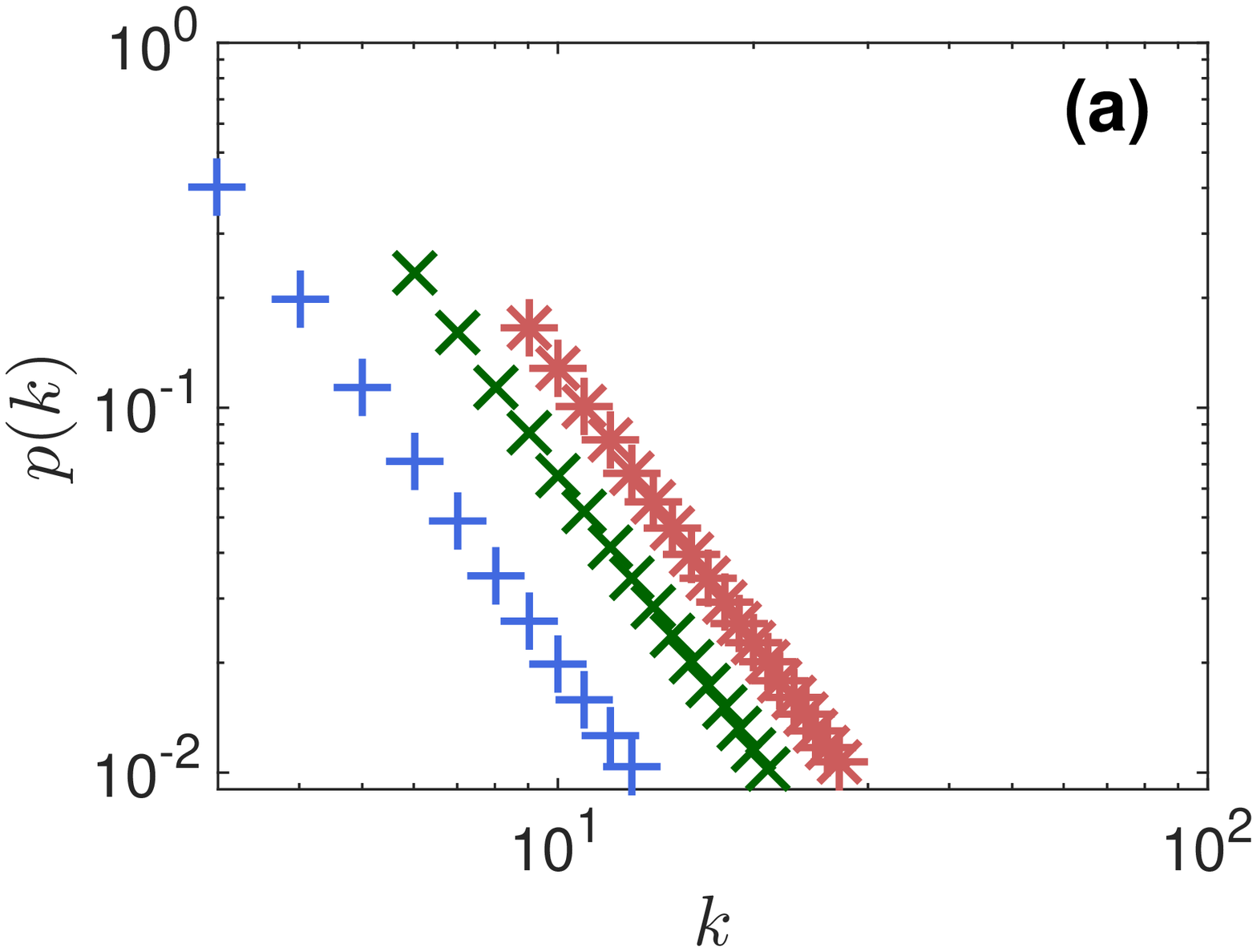}
\includegraphics[width=0.7\columnwidth]{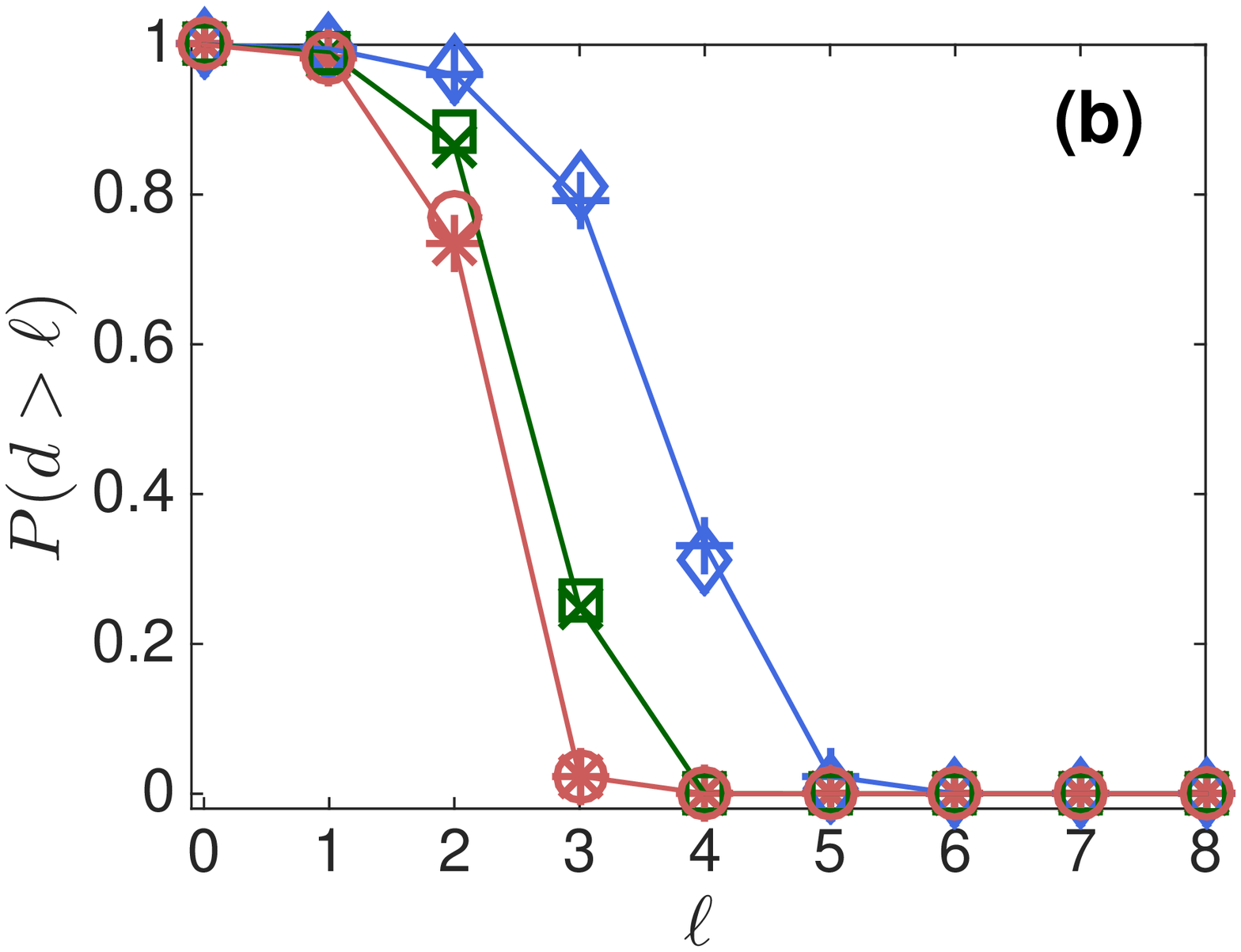}
\caption{
(Color online)
(a) The degree distributions of three networks of size $N=1000$,
where $p(k)$ was drawn from power-law distributions
with 
$\gamma=2.5$ and
lower cutoffs at
$k_{min}=2$, $5$ and $8$
(+, $\times$ and $\ast$, 
respectively). 
The upper cutoffs, $k_{max}$  
were set such that 
$p(k_{max})=10/N$.
The results were obtained from numerical simulations,
averaging over $50$ graph instances.
(b)
The tail distributions
$P(d>\ell)$,
obtained from the recursion equations
($\Diamond$, $\square$ and $\bigcirc$, 
respectively), 
and from numerical simulations
(+, $\times$ and $\ast$, 
respectively),
for the three networks described above.
It is observed that as the lower cutoff,
$k_{min}$, is increased, the mean distance
decreases.
}
\label{fig:sf}
\end{center}
\end{figure}

In Fig. 
\ref{fig:sf}(a)
we present the degree distributions of three scale-free network 
ensembles with $N=1000$ nodes and $\gamma=2.5$. 
The lower cutoffs of the degree distributions of these networks
are given by
$k_{min}=2$, $5$ and $8$, respectively.
In each one of these three ensembles, 
the upper cutoff, $k_{max}$ was chosen such that 
$p(k_{max}) \simeq 0.01$,
which means that in a network of 1000 nodes there will
be on average about $10$ nodes with degree $k_{max}$.
In Fig. 
\ref{fig:sf}(b)
we present the tail distribution
$P(d>\ell)$
for a scale free network with the degree
distributions shown in Fig.
\ref{fig:sf}(a).
The analytical results are in very good agreement 
with the numerical simulations.

In the asymptotic limit, where
$k_{max} \rightarrow \infty$,
the power-law distribution satisfies
$\langle k \rangle = \zeta(\gamma -1,k_{min})/\zeta(\gamma,k_{min})$
and
$\langle k^2 \rangle = \zeta(\gamma -2,k_{min})/\zeta(\gamma,k_{min})$.
Plugging the power-law degree distribution
(\ref{eq:power_law})
into Eqs.
(\ref{eq:genfunc})
and
(\ref{eq:cavgenfunc})
we obtain that

\begin{equation}
G_0(x) = 
\frac{\Phi(x,\gamma,k_{min})}{\zeta(\gamma,k_{min})} x^{k_{min}}
\label{eq:genfuncpl}
\end{equation}

\noindent
and

\begin{equation}
G_1(x) = 
\frac{\Phi(x,\gamma-1,k_{min})}{\zeta(\gamma-1,k_{min})} x^{k_{min}-1},
\label{eq:genfuncplc}
\end{equation}

\noindent
where 
$\Phi(x,\gamma,k)$ 
is the Lerch transcendent
\cite{Gradshteyn2000}.
Evaluating 
$\ell_{\rm mode}$
for a network with a power-law degree distribution
using Eq.
(\ref{eq:mode})
we obtain

\begin{equation}
\ell_{mode} = 
\frac{\ln N}
{\ln \left( \frac{\langle k^2 \rangle 
- \langle k \rangle}{\langle k \rangle}\right)} +{\cal O}(1).
\label{eq:modepowerlaw}
\end{equation}

\noindent
Note that in scale free networks characterized by 
$2<\gamma<3$, 
the value 
of the second moment
$\langle k^2\rangle$ 
is dominated
by the upper cutoff,
$k_{max}$.
As long as 
$k_{max}$
is kept finite, 
$\ell_{mode}$
will depend on this upper cutoff. 
On the other hand, in case that
$k_{max} = N-1$,
then for 
$\gamma=3$ 
one obtains that
$(\langle k^2 \rangle- \langle k \rangle)/ \langle k \rangle$
diverges logarithmically with $N$.
As a result,
$\ell_{mode} \sim \ln N / \ln\ln N$
for large $N$. 
For 
$2 < \gamma < 3$ 
one obtains that
$(\langle k^2 \rangle- \langle k \rangle)/ \langle k \rangle 
\sim (N-1)^{3-\gamma}$,
entailing that 
$\ell_{mode} = {\cal O}(1)$.

The mean distance between nodes in scale free networks was studied
in Ref.
\cite{Cohen2003}.
Using an analytical argument it was shown that 
scale free networks with degree distribution of the form 
$p(k) \sim k^{-\gamma}$
are ultrasmall, namely exhibit a mean distance
which scales like 
$\langle \ell \rangle \sim \ln \ln N$
for $2 < \gamma < 3$.
For $\gamma = 3$ it was shown that the mean distance scales like
$\langle \ell \rangle \sim \ln N/ \ln \ln N$, 
while for $\gamma > 3$
it coincides with the common scaling of small world networks, 
namely 
$\langle \ell \rangle \sim \ln N$.
As of now, our approach does not yield a closed form
expression for the mean and thus we cannot provide a
conclusive result for its scaling with $N$. 
We do see that the scaling of the mode 
of the DSPL coincides with
the scaling predicted
for the mean of the DSPL in Ref.
\cite{Cohen2003}
for $\gamma > 3$.
In the range 
$2 < \gamma < 3$
we find that the mode is of order $1$, namely
independent of $N$, which is even shorter than
$\ln \ln N$.
This is consistent with the ultrasmall scaling of the
mean, reported in Ref.
\cite{Cohen2003},
since the mode is expected to be smaller than the mean
and less sensitive to extreme values.

\section{Summary and Discussion}

We presented a theoretical framework for the calculation of the 
distributions of shortest path lengths 
between random pairs of nodes
in configuration model networks.
This framework, which is based on recursion equations derived
using the cavity approach, provides analytical results for the
distribution of shortest path lengths.
We used the recursion equations to study a broad class of configuration
model networks, 
with degree distributions that follow the
degenerate, binomial and power-law distributions.
The results were shown to be in good agreement 
with numerical simulations. 
The mean, mode and variance of the distribution
of shortest path lengths were also evaluated
and expressed in terms of moments of the degree distribution,
illuminating the important connection between the two distributions.
The DSPL is of great relevance to transport processes on networks
such as information flow and epidemic spreading. 
For example, an epidemic tends to spread outwards from the 
node where it was initiated.
As time proceeds, it may reach nodes in shells farther away
from the initial node and increases the fraction of infected nodes in
the inner shells.
Therefore, the number of nodes in each shell and their
connectivity affect the
rate and efficiency in which the epidemic progresses in the population
\cite{Shao2015}.

The approach presented in this paper is aimed at the calculation
of the entire distribution of distances between pairs of nodes
in configuration model networks. In general, it does not provide a closed
form expression for the DSPL but a set of recursion equations which can
be evaluated for a given network size and a given degree distribution. 
As a result, it is difficult to obtain a closed form expression for the mean
distance, except for special cases such as the regular graph. 
In fact, for the regular graph, our result for
the mean distance coincides with the exact result presented in Ref.
\cite{Hofstad2005}.
Regarding the mode of the DSPL, we do manage to obtain an
analytical expression in the general case.
The mode turns out to be more amenable to analysis than the
mean because it can be determined by a local criterion.
For degree distributions with a finite second moment,
the mean and the mode tend to scale in a similar fashion.
However, in the case of scale free networks, the mean and the
mode may scale differently.
This is related to the fact that
in scale free networks with $2 < \gamma < 3$,
the second moment of the degree distribution,
$ \langle k^2 \rangle$, 
diverges in
the infinite system limit.
The second moment appears in the equations for 
the mean distance and for the mode,
thus calling for a special care
in scale free networks.
The mode is less sensitive to extreme values and therefore is
expected to be smaller.
We find that for $2 < \gamma < 3$ the mode
is of order $1$, namely does not scale with the network size.
Lacking a closed form expression for the mean,
we cannot provide a conclusive
result for the scaling of the mean with the system size.
This is an important issue which deserves further research. 

M.N. is grateful to the Azrieli Foundation for the award of an Azrieli Fellowship.

\newpage

\end{document}